\documentclass{aa}  

\usepackage{graphicx}
\usepackage{txfonts}
\usepackage{xcolor}
\begin{document}

   \title{Feasibility of ultra-high-energy cosmic ray backtracking \\through sparse local measurements of the Galactic magnetic field.}

   \author{Stylianos Romanopoulos\inst{1,2},
          Michalis Mastorakis\inst{1,2}
          \and 
          Vasiliki Pavlidou\inst{1,2}
          }

   \institute{Department of Physics \& Institute of Theoretical and Computational Physics, University of Crete, GR-70013, Heraklion, Greece
   \email{sromanop@physics.uoc.gr, mmastorakis@physics.uoc.gr, pavlidou@physics.uoc.gr}
         \and
             Institute of Astrophysics, Foundation for Research and Technology-Hellas, Vasilika Vouton, GR-70013 Heraklion, Greece
             }


 	\authorrunning{Romanopoulos et al. }
	   \titlerunning{Feasibility of UHECR backtracking through sparse, local  GMF measurements} 
  \abstract
   {Planned and ongoing campaigns for the acquisition of high-quality local measurements of the Galactic magnetic field (GMF) at interstellar cloud locations have generated intense interest in the use of such measurements to accurately backtrack Ultra High-Energy Cosmic Rays (UHECR) through the Milky Way,  a crucial aspect of charge-particle astronomy. However, the inherent sparsity of these measurements raises concerns regarding the feasibility of this approach. }
   {We assessed the achievable accuracy of UHECR backtracking using mock sparse local GMF data derived from the Jansson \& Farrar 2012 (JF12) GMF model and mock UHECR events.}
   {We created mock UHECR datasets that trace back within a $3^{\circ}$ angular range from the galaxy M82 (a hypothesized UHECR source), and we investigated the impact on such backtracking attempts of varying GMF measurement sparsity and of varying GMF strength, which we emulated by rescaling the strength of the ordered components of the JF12 model. }
   {We found that: (a) for an average GMF strength of $1\mu G$, satisfactory backtracking results for magnetic rigidities of $10^{20}$ eV can be obtained even with very sparse measurements ($ \sim 1600$ pc); (b)  when the average GMF strength is significantly increased ($\sim$ factor of 10) the accuracy of backtracking breaks down at measurement spacings of 400 pc. }
   {These findings emphasize on one hand that sparsity is not an automatic deal-breaker for the utility of local GMF measurements in UHECR backtracking. On the other hand, we also confirm that important challenges remain on the path from sparse local GMF measurements to precise charge-particle astronomy, especially in directions of high-strength ordered magnetic fields. This underscores the importance of using all available complementary magnetic field measurements and sophisticated reconstruction techniques 
   to enable accurate backtracking of UHECR.}


   \maketitle
%

\section{Introduction}

With energies $\gtrsim 10^{18}$ eV, some exceeding $10^{20}$ eV, ultra-high-energy cosmic rays (UHECRs) are the most energetic particles in the Universe. Their study is inherently challenging due to their electric charge, which causes them to be deflected by both extragalactic and Galactic magnetic fields (EGMF and GMF). The magnitude of these deflections decreases with increasing  particle rigidity $E/Z$, and increases with the strength of the traversed magnetic field $B$. Despite systematic efforts over the past decades to constrain the EGMF (e.g., \citealp{Angela95, Beck01, Globus08, Neronov10, Dolag11, Pshirkov16}), probe and model the GMF (e.g., \citealp{Manchester72,SNK80,LyneSmith89,Han94, Han06,Noutsos08,Ferriere09,Sun10,Jaffe10,Mao10,VanEck11,Pshirkov11,JF12,Sun15,Planck16, UF23}),  and account for UHECR deflections in such models (e.g., \citealp{Stanev97, AugerDipole, Globus23, Bister24, TA24}), uncertainties in the UHECR composition, the UHECR energy scale, and the structure and strength of the EGMF and the GMF have thus far hindered the full realization of charged-particle astronomy. A significant aspect the problem lies in the nature of GMF observables, which are typically integrated along the line of sight, with  the inversion problem being highly complex and involving significant degeneracies. 

In recent years, advancements in observational and analysis techniques have enabled high-quality {\em local} measurements of the GMF at the locations of interstellar clouds. Surveys such as PASIPHAE \citep{Pasiphae, WALOP-south1, WALOP-south2} and SOUTH POL \citep{Southpol} employ optopolarimetry of stars to provide valuable data on the GMF, based on dichroic absorption of starlight by dust grains in clouds aligned with the magnetic field that permeates them. In combination with stellar distances from Gaia \citep{Gaia1, GaiaDR3}, a tomographic decomposition of the GMF signal can be performed, yielding local measurements in individual clouds \citep{Panopoulou, Pelgrims, Pelgrims-map}. Alternatively, the combination of density and velocity measurements in interstellar clouds can also provide local GMF measurements \citep{Tritsis-method, Tritsis:2018drs}. The prospect of the wide applications of such techniques and the delivery of a large number of local, accurate GMF measurements has renewed interest in the possibility of using them to correct UHECR deflections. However, these measurements are inherently sparse, raising concerns about the feasibility of their use to accurately backtrack UHECRs through the Galaxy. 

In this paper, our aim is to investigate how sparse is "too sparse" for such an endeavor. In particular, we aim to assess the accuracy of backtracking UHECRs through a GMF that is sampled with increasing sparsity, and with uncertainties typical of what is expected by local-GMF-measurement techniques \citep{ST2, ST, Tritsis-method}. In particular, we aim to assess the feasibility of a scenario where backtracking targets a set of events originating from a single source. These events will not only be dispersed due to deflections of their paths in the turbulent component of the GMF, but they will also be systematically displaced away from the source due to the impact of the large-scale ordered component of the GMF. We therefore investigate whether the backtracking we envision will restore their locations not only to a smaller angular distance from the source, but also redistribute them so that the barycenter of their arrival locations is centered, within uncertainties, on the source. 

The answer to this question will be dependent not only on the sparsity of the sampling, but also the location of the source on the sky. There are several reasons for this. First, for sources close to the Galactic plane, only the most nearby clouds can be probed through optopolarimetry, due to heavy extinction suffered by stars at larger distances. Second, particles originating in sources towards the Galactic anticenter have to traverse over 20kpc of Galaxy in order to reach the Earth, including parts where Gaia distances are very uncertain or non-existent \citep{GaiaDistances}, and hence cloud localization is poor or impossible. Third, the GMF itself varies across the Milky Way, so the deflections to be corrected will vary strongly in different directions \citep{Globus23}. Finally, some locations on the sky are of high interest to the UHECR community, due to the existence there of prominent UHECR hotspots (e.g., \citealp{Abbasi,AugerCenA, TA24}), or astrophysical systems hypothesized to be UHECR sources due to their proximity and/or nature. 

Here, we choose to test the feasibility of UHECR backtracking through sparse local measurements for a specific, likely,  favorable scenario: that the nearby starburst galaxy M82 is a UHECR source, producing a dominant fraction of events in its neighborhood. The scenario is likely because (a) of the hinted correlation between starburst galaxies and the arrival direction of the highest-energy cosmic rays \citep{Auger-Starburst}; (b) of the fact that M82, at a distance between 3 and 4 Mpc, is the most nearby starburst galaxy; (c) of the existence of a prominent UHECR hotspot nearby \citep{Abbasi,TA24}.  The likelihood of M82 being the source of the TA hotspot has also been emphasized by \citet{Supergalactic}. The scenario is favorable because of the location of M82 in the sky (high Galactic latitude, towards the Galactic anticenter), minimizing the distance within the GMF a particle has to traverse in order to reach the Earth, and ensuring that Gaia distances with reasonable uncertainties will exist in this direction to localize most clouds where GMF measurements can be obtained. 

As our "ground truth" of the GMF to be sampled through local measurements we  use a model based on  \citet{JF12} (hereafter JF12), with the ordered component of the magnetic field rescaled by some factor. In this way, we wish to test the feasibility of backtracking through sparsely sampled GMFs of varying strength.  We do not make any attempt to interpolate the GMF between measurements or reconstruct features of the GMF; we simply use the nearest measurement of the GMF for each step in the backtracking. For the impact of sophisticated reconstruction techniques on backtracking accuracy see \citet{Tsouros1, Tsouros2}.

For this study we  assume all CR to be protons, and to follow the energy distribution of 74 Telescope Array hotspot events the majority of which have been hypothesized to originate from M82 \citep{Supergalactic}. For a discussion of  composition uncertainties, and of whether backtracking can at the same time constrain the location and composition of a source assuming a strong enough hotspot exists, see \citet{Tsouros3}

This paper is organized as follows. We describe the GMF model and sampling assumptions we used in \S \ref{magnetic}. We present the implementation of backtracking for a single event through the mock magnetic field data of \S \ref{magnetic} in \S \ref{backtracking}. In  \S \ref{mockevents}, we discuss the procedure through which we generate the mock datasets to be backtracked. We present our results in \S \ref{results}, and we discuss our conclusions in \S \ref{conclusions}. 

\section{GMF model and sampling}\label{magnetic}

We based our assumed "ground truth" GMF on the model of JF12. Given that \citet{Tritsis:2018drs} have found that, in the general direction of M82, GMF models may be underestimating significantly the GMF strength, we allowed the strength of the ordered magnetic field in JF12 to vary. In particular, we rescaled the ordered component of the field by a factor $f$, and we examined three values of $f$: $f=1$, $f=3$, and $f=10$. If the results of \citet{Tritsis:2018drs} are typical for this region, the actual value of the ordered field is likely bracketed by $f=1$ and $f=10$. 

 The magnetic field can be measured locally in interstellar clouds within the Galaxy, but these clouds are not close to one another. The distance between them varies.
To simulate this reality of sparse measurements,  as a zeroth order of approximation, and for simplicity and clarity in our analysis, we  assumed that the sampling is uniform, with a certain linear spacing between measurements, which we varied. Determining the dependence of the backtracking accuracy on this spacing is the primary aim of this paper.
In practice, we divided the Galaxy into cubes of side length L and within each cube we assumed a constant magnetic field value and direction that was found by calculating these quantities at the center of the cube using the "ground truth" GMF. We controlled the sampling of the magnetic field values (the number of cubes), by changing the value of L.

We also assumed that all three components of the magnetic field are sampled. This assumption is unlikely to be realistic for the large majority of local measurements. The reason is that most local measurements of the GMF are expected to be obtained through stellar optopolarimetry, which only probes the plane-of-sky (POS) component of the GMF.  However, the effect of this assumption on our results is expected to be small, for two reasons. First, because as long as deflections of UHECRs remain small (few degrees), their velocity is mostly parallel to the line of sight, and the sensitivity to the line-of-sight (LOS) component of the magnetic field is  small. Second, because in practice POS local measurements can be supplemented by integral measures of the LOS component of the magnetic field (e.g. from Faraday rotation measurements of extragalactic sources) in the same direction. \citet{Tsouros2} have explicitly shown that even an approximate, average measurement of the LOS component of the magnetic field is sufficient as a supplement to POS measurements to yield highly accurate backtrackings, given the small sensitivity of UHECR deflections to LOS fields. 

Finally, we assumed that observational uncertainties are heavily dominated by uncertainties in the strength, rather than the direction, of the magnetic field. This is indeed a realistic assumption for local GMF measurements based on optopolarimetry, as the strength of the magnetic field is obtained through estimates of the dispersion of polarization directions (e.g. \citealp{ST2, ST}), while the direction is obtained from the average of these directions, which is recoverable with much higher accuracy. We examined two cases: uncertainty of 25\% and 50\% in the value of the magnetic field strength in each measurement. In practice, we implemented measurement error in the following way. After calculating the GMF at the center of each cube in our grid, we resampled a new "observed" value of its strength from a  Gaussian distribution with mean that of the "ground truth" and a standard deviation of 0.25 or 0.5 times that value. As a control, in order to be able to assess the effect of measurement sparsity alone on the quality of reconstruction, we also repeated the backtracking experiment using GMF "observations" free of error. 


\section{Implementation of backtracking}\label{backtracking}


The equation of motion of an ultrarelativistic nucleus with charge $q = Ze$ (where $Z$ is the atomic number and $e$ is the charge of an electron) moving through a magnetic field $\vec{B}$ and an electric field $\vec{E}$ is given by:
\begin{equation}\label{general eom}
\dfrac{d}{dt}\left(m\gamma\vec{v}\right) = q\left(\vec{E} + \vec{v}\times\vec{B}\right) 
\end{equation}
where $\gamma$ is the Lorentz factor. Assuming $\vec{E} \approx 0$ in interstellar space, $d\gamma/dt \approx 0$ and $\vec{v}\approx c\hat{v}$. Additionally, for ultrarelativistic particles $m\gamma\vec{v}\approx (E/c)\hat{v}$ where $E$ is the particle energy, so the equation of motion becomes 
\begin{equation}\label{general eom2}
\frac{E}{c}\dfrac{d}{dt} \hat{v}  \approx  Zec\hat{v}\times\vec{B}
\end{equation}
which discretizes to 
\begin{equation}\label{velnumint}
\hat{v}_{\text{prev}} = \hat{v}_{\text{now}} - \dfrac{Zec^{2}}{E}\left(\hat{v}\times\vec{B}\right)\delta t.
\end{equation}
Similarly, for the position vector of the particle $\vec{r}$ we can write the discretized equation 
\begin{equation}\label{posnumint}
\vec{r}_{\text{prev}} = \vec{r}_{\text{now}} - \vec{v}\delta t = \vec{r}_{\text{now}} - \hat{v}c\delta t.
\end{equation}
Our backtracking code uses equations (\ref{velnumint}) and (\ref{posnumint}) to numerically calculate the position and velocity of a UHECR during the backtracking process.

In this paper, we assume that cosmic rays consist solely of protons ($Z = 1$). 
The starting position for the backtracking of all cosmic rays is set to $\vec{r} = (-8.5, 0, 0)$ kpc in a Cartesian system centered on the Galactic center, as the backtracking process starts at the location of the Earth. The velocity unit vector (i.e., the direction of velocity) is determined from the Galactic coordinates of a reported event, $(l, b)$ as follows:
\begin{equation} \label{velocity from galactic coords}
            v_{z} = \sin(b), \ v_{x} = \cos(b)\cos(l), \\
            v_{y} = \cos(b)\sin(l)\,.
    \end{equation}
Conversely, Galactic coordinates of an event's updated (backtracked) position are determined from the velocities as follows:
\begin{equation} \label{galactic coords from velocity}
            b = \arcsin(v_{z}), \\
            l = sign(v_{y})\arccos(\frac{v_{x}}{\cos(b)}),
    \end{equation}
and if the resulting $l$ is negative, we increase it by $2\pi$. We assume the source to be distant enough so that the size of the Milky Way is insignificant compared to the distance between Earth and source, so the direction of the velocity of a particle before it enters the Milky Way magnetic field coincides with the arrival direction that would be recorded for a particle from the same source if no Galactic magnetic field existed. 

Additionally, we neglect the intergalactic magnetic field and assume that cosmic rays propagate as free particles until they enter the Galaxy. Note that this is a different approach than the assumption of \citet{Supergalactic}, who have assumed that the largest fraction of deflections causing the spreading out of events from particular sources, and, specifically, from M82, is due to the intergalactic magnetic field, with the Galactic field imparting small deflections by comparison. The \citet{Supergalactic} assumption is reasonable given the small deflections imparted by the GMF on protons according to, e.g., the JF12 model, compared to the angular spread of the hotspot. However, if either the composition is heavier or, as in the case we examine here, the GMF in that direction is stronger than models predict (e.g. \citealp{Tritsis:2018drs}), the GMF deflections for such a nearby source could very well be dominant. 

The angular distance of a backtracked event from the source is given by:
\begin{equation} \label{angular position}
            \theta = \arccos \left[\sin b_{cr} \sin b_{M82} + \cos b_{cr} \cos b_{M82} \cos\left(l_{cr} - l_{M82}\right) \right],
\end{equation}
where $l$ and $b$ represent the longitude and latitude in Glactic coordinates, respectively, and the subscript cr stands for "cosmic ray". 

\section{Generation of mock event datasets}\label{mockevents}

In this section, we describe the process through which we produced  mock event datasets originating from our putative source (M82), to examine their behavior under backtracking through sparse measurements of the GMF. To do so, we ensured that the mock events backtrack to our desired location (within 3$^\circ$ from the position of M82 in the sky) if the "ground truth" magnetic field (see \S \ref{magnetic}) is used for the backtracking. We chose to generate events with this finite spread around the source, to account on the one hand for uncertainties in the determination of arrival directions from cosmic ray observatories (on the order of a degree) and in the CR energy (around 10\%), and on the other hand for diffusive deflections by intergalactic fields (unknown, but for a source as close as M82 it is plausible to assume that these deflections will be  of that order of magnitude). We did not apply any energy dependence to this scatter (experimental + extragalactic) about M82. 

\begin{figure}[h]
    \centering
    \includegraphics[width = 1.05\columnwidth]{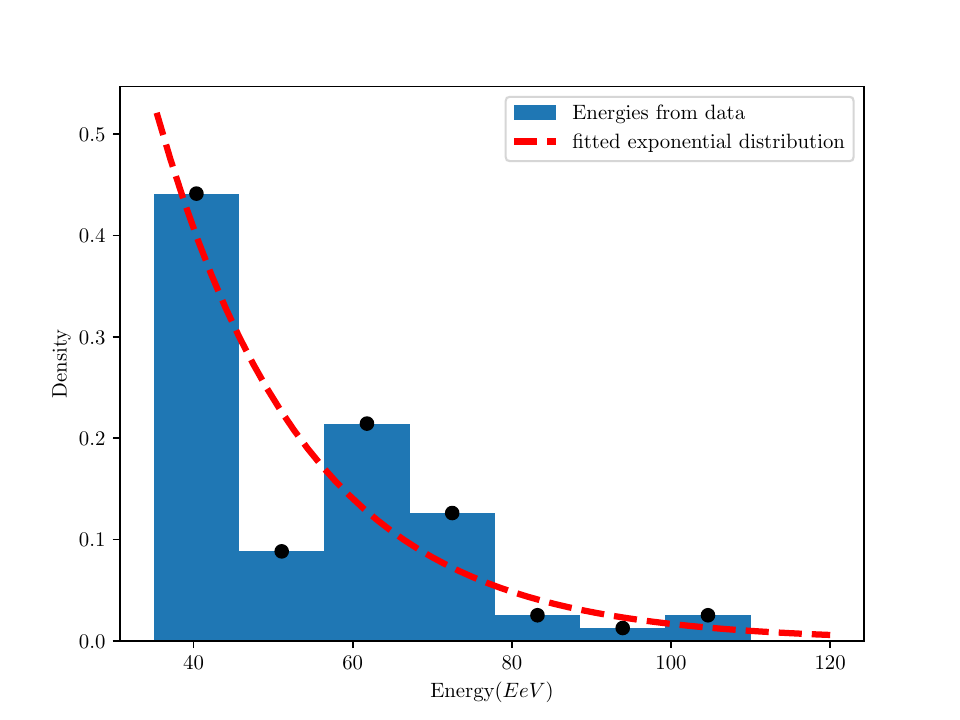}
    \caption{Distribution (probability density) of the energies of the 74 real Telescope Array events with high likelihood of originating from the data. The red dashed line shows the shifted exponential distribution fit of Equation \ref{shifted exponential distribution}, used to derive energies of the simulated event datasets.}
    \label{fig:energies histogram}
\end{figure}

We also assumed a "ground truth" for the energy spectrum of the particles it produces. As detailed above, we assumed that M82 is a source of UHECRs with an energy spectrum that of the 74 UHECRs studied by \cite{Supergalactic} as likely to have originated from M82. Rather than bootstrap from this small dataset, we chose to fit the distribution of energies reported in \cite{Supergalactic} for these events  with  a shifted exponential probability density function (PDF) of the form \begin{equation}\label{shifted exponential distribution}
    f(E ; \lambda) = \lambda \exp \left[-\lambda(E- k)\right],
\end{equation} where $\lambda=0.53 {\rm EeV}^{-1}$ is the parameter of the distribution and $k=34.96 {\rm EeV}$ the shift in the position of the distribution. The exponentially suppressed (rather than power-law) form of the energy spectrum is consistent with expectations for events at the edge of the UHECR spectrum. In Fig. \ref{fig:energies histogram} we confirm that this exponential fit is a good approximation for the distribution of energies. 

 For each simulated UHECR, we need three quantities: (1) its energy $E$; (2) the longitude $l$ characterising its arrival direction; and (3) the latitude $b$ characterising its arrival direction. All other parameters for our cosmic rays are either fixed (for example, their starting position in the Galaxy is that of the Earth's; their atomic number  is $Z = 1$), or they can be derived from these 3 quantities [for example, the magnitude of their velocity is $\approx c$ and its direction can be found using the latitude and longitude from eqs. (\ref{velocity from galactic coords})]. 
 
We produced simulated events using a Monte Carlo approach, as follows. We first drew a value for the energy of a mock event from the energy distribution of Eq.~(\ref{shifted exponential distribution}). We then drew a value for each of $l$ and $b$, representing a random location on the sky. We drew $l$ from a uniform distribution in the range $[0, 2 \pi]$. We drew $\cos b$ from a uniform distribution in the range $[-1, 1]$ (once the cosine is determined, the value of $b$ is obtained by taking the arccos). We then backtracked the cosmic ray through the "ground truth" GMF, and  checked whether its post-backtracking position (i.e., its inverted velocity vector before it entered the Milky Way) pointed within $3^{\circ}$ from M82. If it did, we kept the event. If it didn't, we draw a different pair of $l$ and $b$ and repeat the process until we identified a particle originating within $3^{\circ}$ from M82. In this way, we ensured that the energy distribution of accepted events was indeed the desired one. 
We repeated the process until each dataset we created (one for each value of $f$ determining the scaling of the ordered component of the GMF) contained 1000 events. 

\begin{figure}[h!]
    \centering
    \includegraphics[width = 1.05\columnwidth]{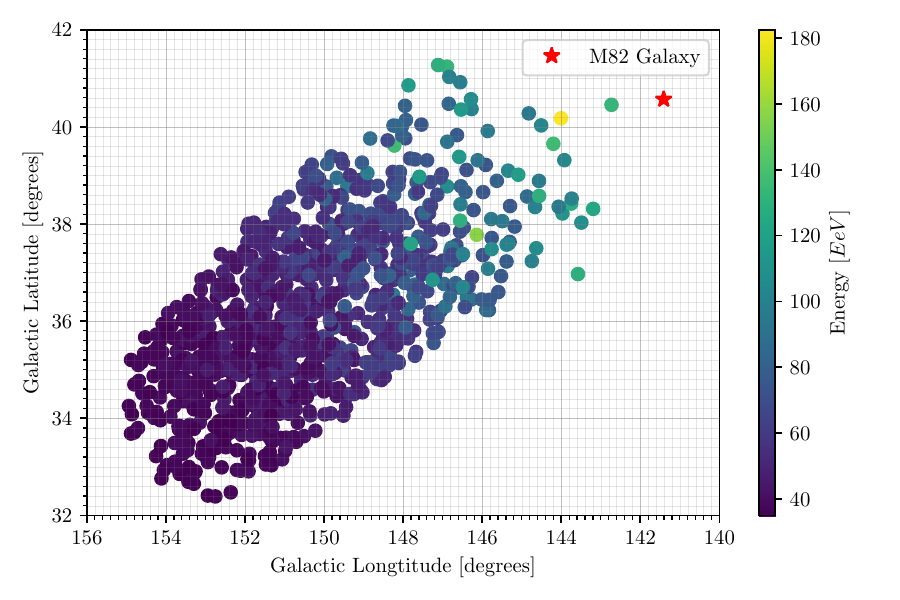}
    \caption{Sky distribution of simulated dataset generated using $f = 3$ as the amplification factor for ordered component of the GMF, before the backtracking process (as they would appear when detected). The colorscale indicates the event energy. M82 is marked by a red star. Deflections are systematically larger for lower energy events. Due to a strong ordered component of the GMF, events are deflected systematically to one side of the source.}
\label{fig:simple_plot_fake_events_3_before}
\end{figure}

\begin{figure}[h!]
    \centering
    \includegraphics[width = 1.05\columnwidth]{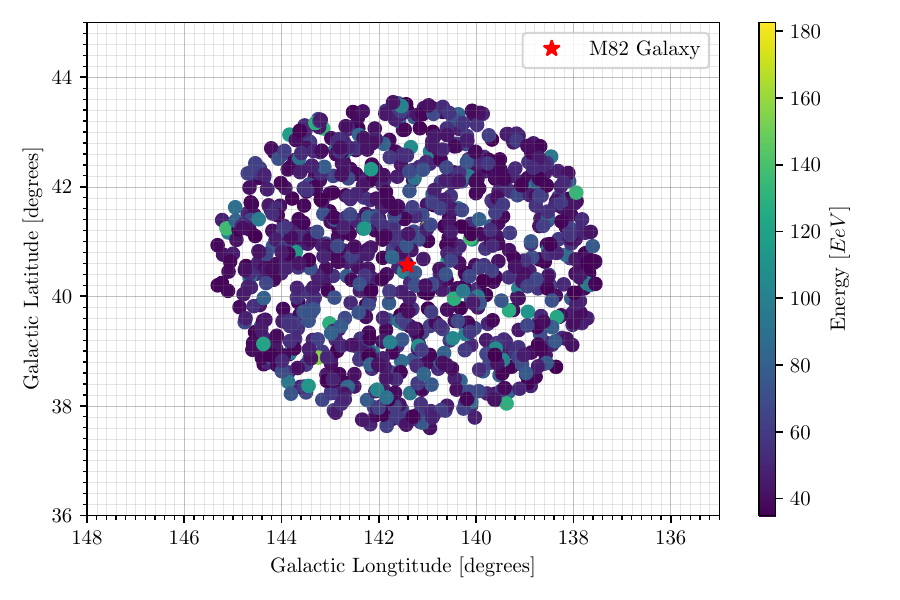}
    \caption{Sky distribution of simulated events generated  $f = 3$ as the amplification factor for ordered component of the GMF, after their backtracking through the ground-truth GMF. Symbols as in Fig.~\ref{fig:simple_plot_fake_events_3_before}.}
\label{fig:simple_plot_fake_events_3_after}
\end{figure}

As a verification of our process, we plot, in Figure     \ref{fig:simple_plot_fake_events_3_before}, the  distribution of simulated arrival directions for events produced for $f=3$ (JF12 GMF model, ordered field rescaled by a factor of 3 upwards). The location of M82 is noted with the red star, while the color bar indicates the CR energy. As expected,  the highest energy events deflect less, while lower-energy events deflect more and are found at greater angular distances from the source. If we backtrack the event set through the "ground truth" GMF with which they were produced, they will, as expected (by design), become uniformly distributed within 3$^\circ$ from M82, without any correlation between event energy and angular distance from the source (see Figure \ref{fig:simple_plot_fake_events_3_after}).

\begin{figure}[h!]
    \centering
    \includegraphics[width = 1.05\columnwidth]{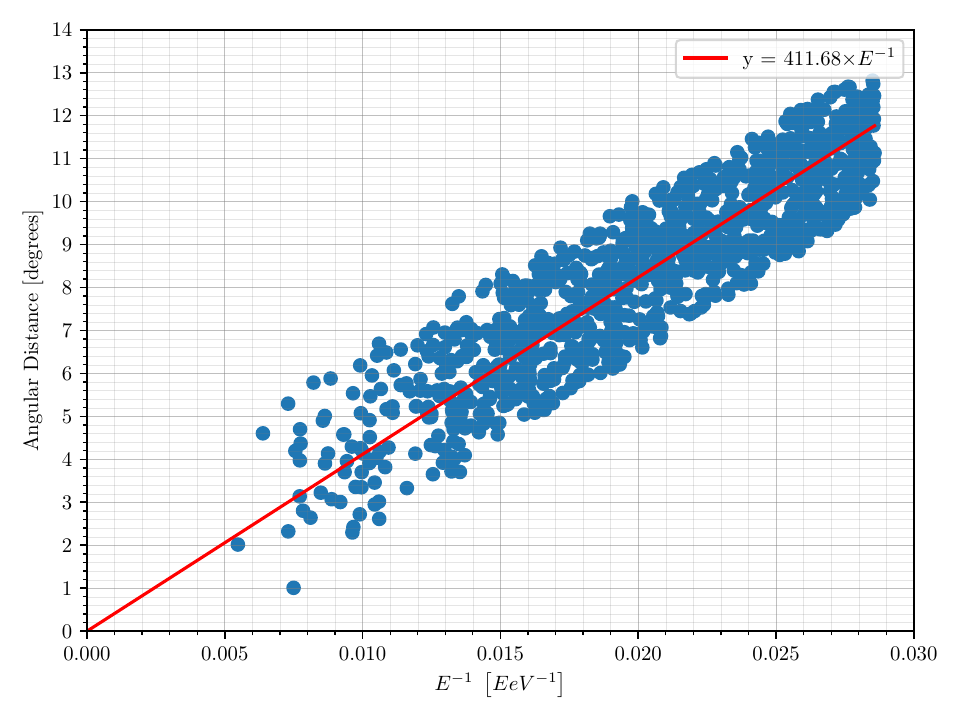}
    \caption{Angular distance from M82 plotted against $E^{-1}$  for the $f = 3$ simulated dataset before the backtracking process (as they would appear when detected). A clear linear correlation with energy is apparent.} 
\label{fig:ang_dist_fake_events_3_before}
\end{figure}

In an ideal future of abundant statistics at these highest energies, and in the favorable scenario of a strong source (where the number of events from a single source would be significantly larger than the background), this dependence of deflection on energy could be used as a first qualitative identification of likely sources, as well as an indicator for the success of any implemented backtracking. We demonstrate these point in Fig.~\ref{fig:ang_dist_fake_events_3_before}, where the angular distance of simulated arrival directions from M82 has been plotted against $E^{-1}$ for each event. As expected, a linear correlation (albeit with significant scatter) is immediately apparent. The extent of the scatter compared to the dynamical range of average deflections for the different energies in our simulated event set is controlled by the relative magnitude of the ordered component of the GMF (which is responsible for the systematic deflections) and the random component of the GMF (and the effect of the extragalactic magnetic field), which is responsible for the random diffusion of arrival directions away from the source. A successful backtracking would erase this correlation. Residual correlation would be an indication of an insufficient magnetic field model / set of measurements used in the backtracking, especially as far as the ordered component is concerned.

\section{Results}\label{results}


We backtracked each of the 3 sets of 1000 simulated events (corresponding to different values of the GMF scaling factor $f$) through the simulated local measurements of the GMF described in \S \ref{magnetic}. We repeated the experiment   for various values of the "slice length" $L$, and for each of the three levels of observational errors we examined for the magnetic field strength (no error; 25\% error; 50\% error). 
After the process of backtracking, we calculated the mean angular distance from M82, $\overline{\theta}$, using  \begin{equation}\label{sample mean}
    \overline{\theta} = \frac{1}{N}\sum_{n = 1}^{N}\theta_{n},
\end{equation} where $N= 1000$, and $\theta_{n}$ is the angular distance from M82 of the n-th CR. 
We also calculated the sample variance for the angular distance $\sigma_{\theta}$, \begin{equation}\label{sample variance}
    \sigma_{\theta} = \sqrt{\frac{1}{N-1}\sum_{n=1}^{N}(\theta_{n} - \overline{\theta})^{2}}.
\end{equation} 

\begin{figure}[h!]
    \centering
    \includegraphics[width = 1.05\columnwidth]{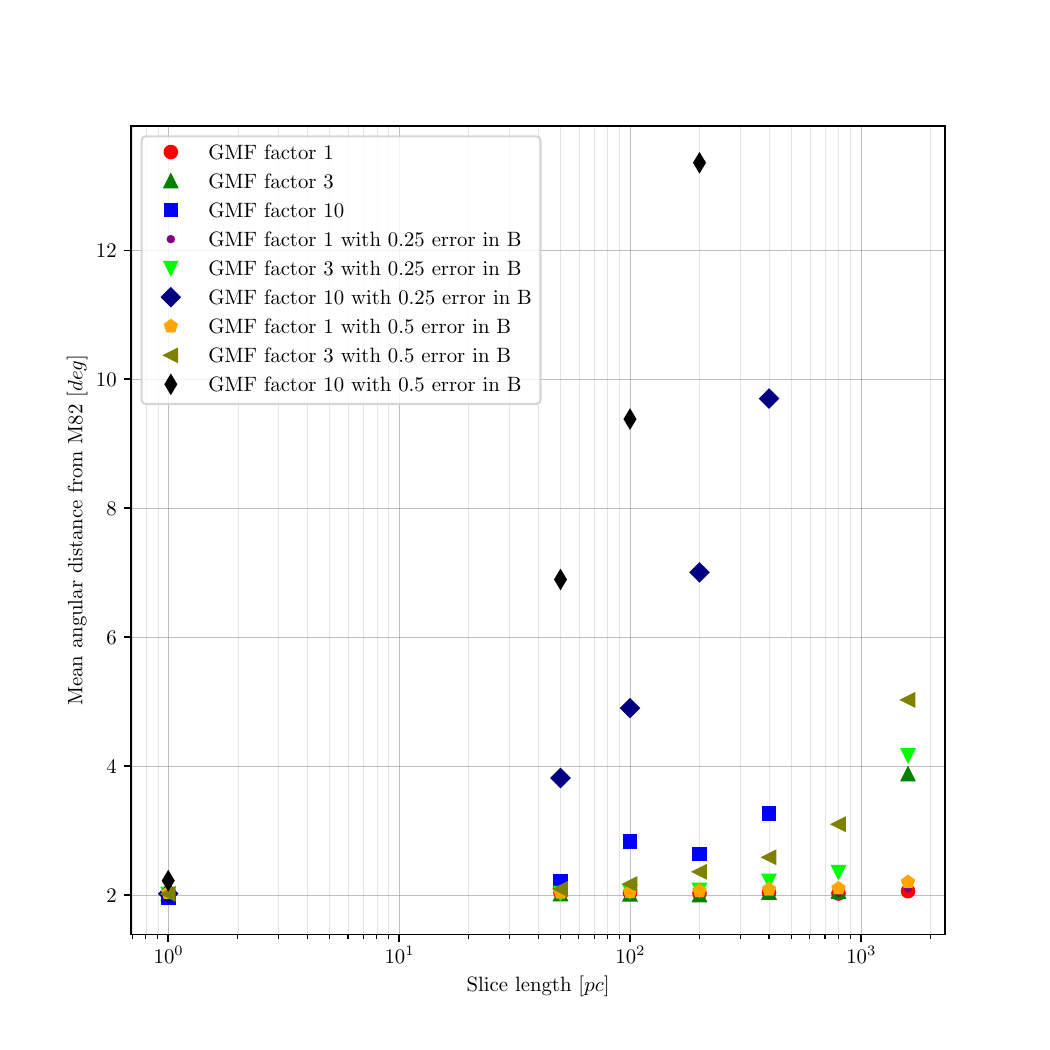}
    \caption{Mean angular position of our simulated events after the backtracking, as a function of the slice length of the cubic grid, $L$, of magnetic field measurements.}
    \label{fig:mean_angular_distance_cubic_grid}
\end{figure}

\begin{figure}[h!]
    \centering
    \includegraphics[width = 1.05\columnwidth]{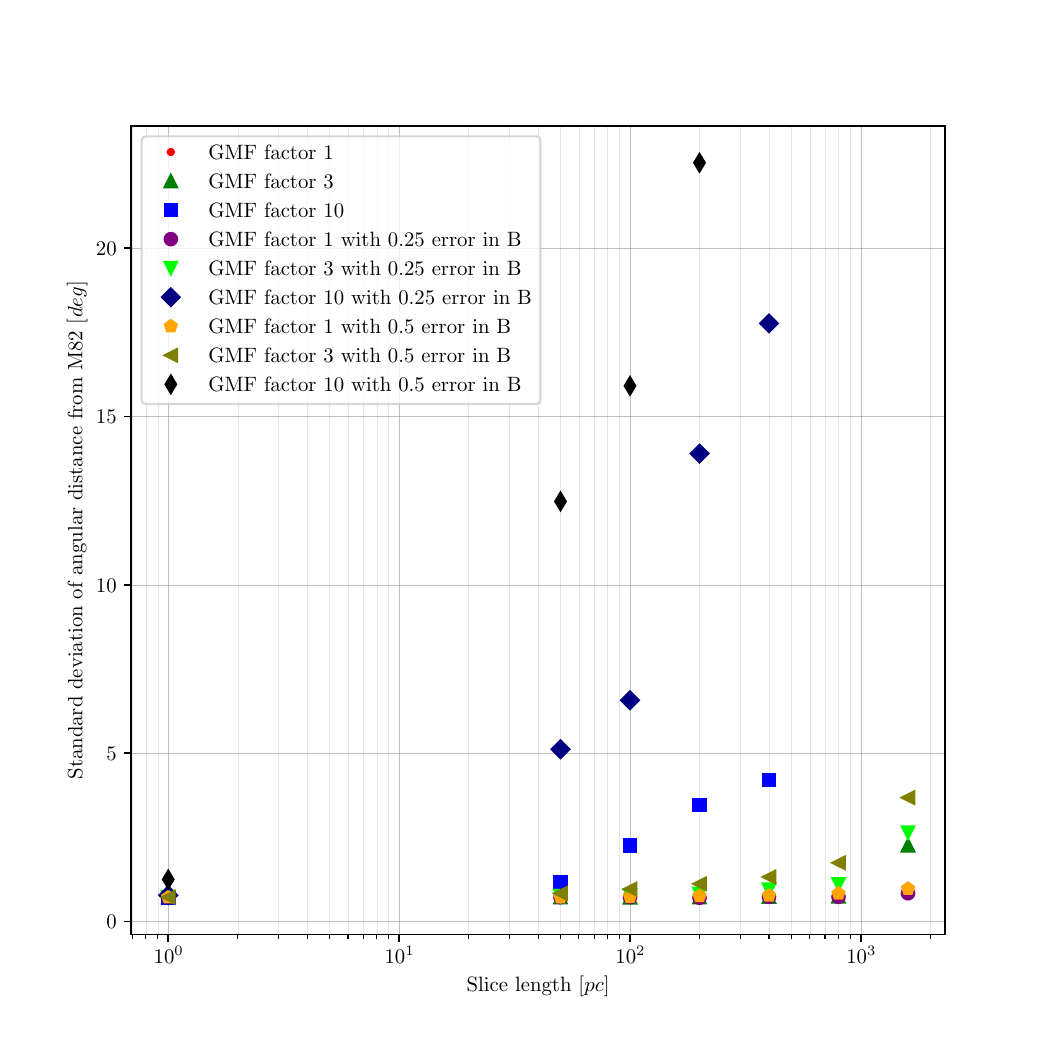}
    \caption{Standard deviation of the angular position of our simulated events after the backtracking, as a function of the slice length of the cubic grid, $L$, of magnetic field measurements. }
    \label{fig:standard_deviation_cubic_grid}
\end{figure}

The full extent of our parameter-study results are shown in Figs.~\ref{fig:mean_angular_distance_cubic_grid}  and \ref{fig:standard_deviation_cubic_grid}, 
In Fig. \ref{fig:mean_angular_distance_cubic_grid} we show the mean angular distance $\overline{\theta}$ of our backtracked events from M82, as a function of the "slice length" $L$. The standard deviation $\sigma_{\theta}$ of angular distances of backtracked events is shown in Fig.~\ref{fig:standard_deviation_cubic_grid}, also 
as a function of $L$. In each case, different symbols correspond to different combinations of GMF scaling factors and errors in the measurement of the magnetic field strength. The "control run" (measurements with unrealistic spacing of 1pc and no measurement error) indicates the residual effect on $\overline{\theta}$ and $\sigma_\theta$ of the scatter we introduced in our "ground truth" arrival directions (which come to $\sim 2^\circ$ and $\sim 1^\circ$, respectively). 

Interestingly, the detrimental effect of sparsity in GMF local measurements strongly depends on the actual strength of the ordered component of the magnetic field in the region of interest (i.e., on our rescaling factor $f$). For an ordered GMF as high as a factor of 3 higher than the one in JF12, $\overline{\theta}$ deteriorates less than a factor of 3, even for extremely sparce sampling ($L>1$ kpc) and high observational errors ($50\%$) in $B$. In contrast, if the ordered GMF component is much stronger than the current models predict ($f=10$), then even with $25\%$ error $\overline{\theta}$ reaches 10 degrees for linear distance between local GMF measurements of $\sim$few hundered parsecs. This could still be useful in establishing the existence of UHECR hotspots on intermediate scales using post-backtracking datasets. This is especially true given that the spread  between events, as quantified by $\sigma_\theta$ is similarly constrained to $\sim 10-20^\circ$ for values of $L$ in the few hundred parsec range. However, locating a source (i.e., a low-energy counterpart for the hotspot) would be more challenging if the ordered GMF is indeed this strong.

\section{Conclusions and Discussion}\label{conclusions}
In this work, we have performed an extensive parameter study examining the combined effect on the accuracy of UHECR backtracking through the GMF back to a suspected source  of: (a) the linear distance between local GMF measurements (sparsity); the level of observational uncertainty in the magnetic field strength (error); and the true strength of the ordered component of the GMF (strength). 

Perhaps surprisingly, we have found that sparsity only has a modest result on the backtracking accuracy if both the GMF error of individual measurements is small and the strength of the GMF itself is not much larger than current models predict. This is a very strong motivation for pursuing local magnetic field measurements. Even sparse local measurements could establish the strength of the ordered component of the GMF. Whether such measurements confirm current model predictions, or further support the deviation that has been indicated by, e.g., 
\cite{Tritsis:2018drs}, they will establish the regime in which we have to operate (in this work's terminology, the likely value of $f$). If the GMF strength turns out to be modest ($f \lesssim 3$), then the strategy should be to obtain as many local measurements as possible (even if only moderate accuracy is achievable), as the tolerance to error is significant. If on the other hand the GMF strength turns out to be higher than that, then emphasis should be placed not only on the number of GMF measurements, but also on their accuracy, as the error tolerance is low. 

We emphasize that in this work no attempt to reconstruct the magnetic field, or interpolate between measurements, has been made. In every backtracking step, we simply adopted the nearest magnetic field measurement as the local value of the magnetic field. The implications are twofold. 
First, in the favorable scenario (many local measurements are feasible, $L$ is small; the GMF ordered component strength is modest, $f\lesssim 3$; good local measurements of the GMF strength are possible, error is small), then UHECR backtracking can be extremely effective, even if no sophisticated reconstruction or modeling of the GMF is performed. Second, in the unfavorable scenario where any one or more of the conditions above do not hold, our results here should only be considered as a lower limit on the quality of UHECR backtracking that can be feasible: there is ample room for improvement using sophisticated reconstruction methods and self-consistent combinations of datasets sensitive to different aspects of the GMF properties.




\section*{Acknowledgments}
This work was supported  by the Hellenic Foundation for Research and Innovation (H.F.R.I.) under the “First Call for H.F.R.I. Research Projects to support Faculty members and Researchers and the procurement of high-cost research equipment grant” (Project 1552 CIRCE). V.P. acknowledges support from the Foundation of Research and Technology - Hellas Synergy Grants Program through project MagMASim, jointly implemented by the Institute of Astrophysics and the Institute of Applied and Computational Mathematics. S.R. acknowledges support from the innovation programme under the Marie Sklodowska-Curie RISE action, grant agreement No 691164 (ASTROSTAT).

%
%
\bibliographystyle{aa}
\bibliography{bibliography}


\end{document}